# Scanning tunnelling spectroscopy of the vortex state in NbSe$_2$ using a superconducting tip

J.G. Rodrigo[*], V. Crespo and S. Vieira

*Laboratorio de Bajas Temperaturas, Departamento de Física de la Materia Condensada,*

*Instituto de Ciencia de Materiales Nicolás Cabrera, Facultad de Ciencias*

*Universidad Autónoma de Madrid, 28049 Madrid, Spain*

**Abstract**

The vortex electronic structure in the multiband superconductor NbSe$_2$ is studied by means of Scanning Tunneling Spectroscopy (STS) using a superconducting tip. The use of a superconducting tip (Pb) as a probe provides an enhancement of the different features related to the DOS of NbSe$_2$ in the tunneling conductance curves. This use allows the observation of rich patterns of electronic states in the conductance images around the vortex cores in a wide range of temperature, as well as the simultaneous acquisition of Josephson current images in the vortex state.

**1. Introduction**

It is widely recognized that scanning tunnelling spectroscopy (STS) is a powerful tool to obtain the local density of states (LDOS) of conducting materials, even at atomic scale. The obtained information is a convolution of the DOS of both electrodes, tip and sample. Since the initial works by Hess et al. [1] on NbSe$_2$, this technique has been used to investigate the electronic density of states in the vortex state in a variety of materials. Improvements in this technique have produced results evolving from the initial observations of the Abrikosov vortex lattice, to detailed conductance maps at selected energies, showing features like electronic bound states at and around the vortex core.

The most detailed results about the symmetry and patterns presented by these bound states (i.e.: six-fold in NbSe$_2$, [1,2] four-fold in YBa$_2$Cu$_3$O$_7$ [3], Bi$_2$Sr$_2$CaCu$_2$O$_{8-d}$ [4] and YNi$_2$B$_2$C [5]) were obtained at low temperatures. There is a great interest in this type of measurements due to the fact that the symmetry of the bound states around the vortex core is related to the symmetry of the order parameter.

Very low temperature is therefore a requisite in order to prevent the thermal smearing of the features associated to the bound states, which typically have an energy width below 0.1 meV. The experiments by Hess et al. [1] and Pan et al. [2] on NbSe$_2$ at 300 mK showed a wealth of patterns associated to the electronic bound states that cannot be obtained at higher temperatures using normal metal tips. .

In the recent years there is also a renewed interest in the topic of multiband superconductivity (MBSC), boosted by the discovery of the superconducting properties of MgB$_2$[6]. Results obtained by different techniques on NbSe$_2$ [7] indicate that this material is also a multiband superconductor. In a previous work [8], we studied the evolution of the distribution of gap values in NbSe$_2$ as a function of temperature. The results of these tunnel spectroscopy measurements, from 300 mK up to the critical temperature (7.2 K), were interpreted as an example of MBSC with small interband scattering.

---

[*] Corresponding author. Tel.: +34 91 497 3800; fax: +34 91 497 3961; e-mail: jose.rodrigo@uam.es

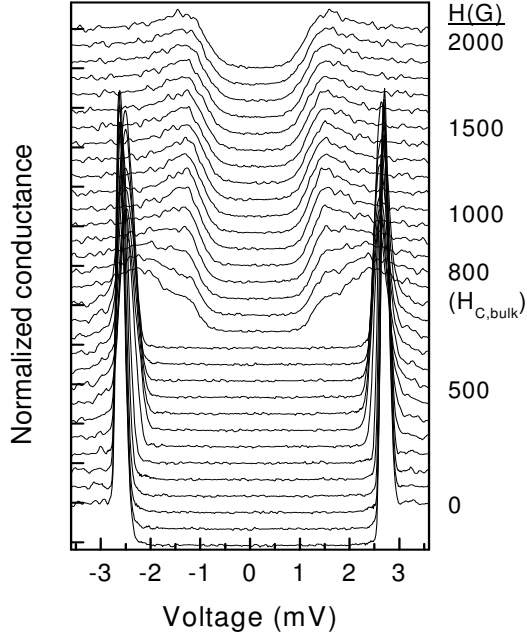

Fig. 1. Pb-Pb conductance curves taken at 300 mK in tunneling regime ($R_N=10M\Omega$) as the magnetic field is varied from 0 to 2000G. This particular nanotip presented a critical field of 9000G.

We were able to observe this behavior by using a superconducting tip (Pb) as counter electrode. This allowed the enhancement of any feature associated with a superconducting gap (due to the convolution of the divergences in the DOS at the gap edges in both electrodes) in all the temperature range.

Therefore, it seems straightforward to investigate the possibilities of using a STM with a superconducting tip to study the electronic bound states around the vortex core at different temperatures. Its evolution up to $T_C$ will add information about the possible variations of the order parameter in the different bands of a MBSC like $NbSe_2$.

Recently, Kohen et al.[9] have used a superconducting tip (Nb) to perform STS on $NbSe_2$ in the vortex state at 2.3K and 4.5K. Their results show variations of the conductance spectra due to the presence of vortices, but the smearing of the curves prevents the observation of the above mentioned bound states and its six-fold symmetry around the vortex core.

In this article we present STS measurements on $NbSe_2$ using a superconducting tip. The electronic bound states and its six-fold pattern can be observed in a large temperature range from 300 mK to 6 K, and for magnetic fields in the range of 1000 G. The in-situ method of preparation of the tip [8,10] produces atomically sharp tips with stable and reproducible superconducting properties, capable of atomic scale topographic resolution. This is of crucial importance in order to obtain high resolution conductance maps, preventing an undesired average of features appearing in the local DOS at the nanoscale [11].

## 2. Experimental

We have used a home built STM, whose sample holder can move in a controlled way millimetric distances in the x and y directions, with nanometric precision. The STM is installed in an Oxford $^3$He refrigerator, with automatic thermal control. Three different materials are located on the sample holder: lead, gold and $NbSe_2$. The lead nanotip is fabricated and characterized on the lead substrate as indicated in a previous work[8]. This process is done at 300 mK, and once we have verified the quality of the superconducting DOS of the tip and its atomic scale resolution by scanning the $NbSe_2$ surface, we proceed to characterize its behavior under magnetic field.

The tip is located on the lead sample in tunneling regime. Well defined superconducting-superconducting spectra are obtained as the magnetic field is increased from zero up to the lead critical field (800G at 300 mK), when the sample becomes normal. At higher fields we obtain normal-superconducting spectra because the nanotip remains superconducting due to its small dimensions compared to the superconducting coherence length and magnetic field penetration depth. Depending on the sharpness of the tip achieved in the fabrication process, the critical field of the nanotip can be as high as 2 T at 300 mK [10]. For the values of the magnetic field used to investigate the vortex states (around 0.1T) the tip presents a superconducting DOS with a well defined gap.

After the characterization of the tip we proceed with the study of $NbSe_2$. The STS measurements consist of 128x128 or 256x256 points topographic images, with an I-V tunneling curve taken at each topographic point. Depending on the number of points of the I-V curve (from 512 to 2048) and the number of topographic points, the acquisition time of the full STS image ranges from 15 to 75 minutes. The use of a software controlled feedback loop in the constant current scanning mode allows faster and more versatile scanning than the usual analog feedback.

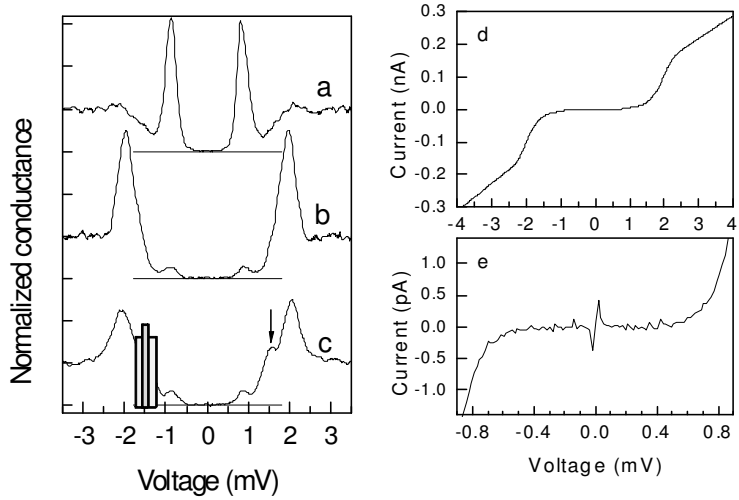

Fig. 2. Pb-NbSe$_2$ conductance curves taken at 300 mK in tunneling regime (R$_N$=10MΩ) for an applied field of 0.1T: (a) at the vortex core, (b) at the mid-point among three vortices, and (c) at an intermediate location (c). The arrow indicates the "extra" peak discussed in the text. The rectangles in curve c indicate the typical ranges of energy where we calculate the local averaged conductances G1 (small range) and G2 (larger range) used to produce the CR and CD images described in the text. Frame (d) shows an I-V curve taken at the mid-point among three vortices. The region close to zero bias is zoomed in frame (e) showing Josephson current.

The scanning range is set depending on the applied magnetic field, in order to allow for a clear imaging of the triangular Abrikosov vortex lattice, ensuring the quality of the results. The bias voltage in the I-V curves is ramped between +5 and -5 mV, to permit the determination of the normal state conductance well above the total gap (tip+sample) in the DOS. These I-V curves are numerically derivated to obtain the conductance curves (G-V) used to produce the conductance images at a given voltage, G(x,y,V). As a routine, small range topographic images are taken in the studied areas to ensure that the superconducting tip keeps its atomic resolution capability.

## 3. Results and discussion

The results obtained during the process of fabrication and characterisation of the superconducting nanotip, both in the absence and in presence of magnetic field, were described in previous articles[8,10]. In fig.1 we present several conductance curves corresponding to a series used to characterize the behaviour of the lead nanotip under magnetic field at 300 mK. These curves, taken on the lead sample, show how a typical S-S spectrum at zero field evolves into a N-S situation once the critical field of the sample is reached. The sharpness of the tip allows the existence of a well defined superconducting DOS in the nanotip for the values of magnetic field that will be used in vortex imaging.

Next, we describe the different features that are present in the conductance spectra obtained on the NbSe$_2$ sample under magnetic field that will lead to the different patterns observed in the conductance images. Fig. 2 shows three curves obtained at different locations on the sample: at the vortex core (a), at the mid-point among three vortices (b), and at an intermediate location (c). As a superconducting tip is used, features in the conductance curves corresponding to the DOS of the sample will appear shifted by $\Delta_{tip}$. Then, it is clear the identification of the sharp peak at 0.8 mV in curve 2a as the one corresponding to the electronic bound state at the vortex core, which as observed by Hess et al. [1] appears at the Fermi level of the sample. Far from the vortices, the superconducting order parameter of the sample will achieve the maximum value. Therefore, the large peak at 2 mV in curve 2b corresponds to $\Delta_{tip}+\Delta_{sample}$. This curve presents a small peak at 0.8 mV, at $E_{Fsample}$, due to the extended bound state at Fermi energy that leads to the well known six-fold star shaped vortex core in NbSe$_2$. A small feature observed at zero bias is due to pair tunneling between tip and sample (see the I-V curve shown in 3d and 3e), and it allows to produce a Josephson spectroscopy map of the vortex state. This Josephson imaging of the vortex state will be discussed elsewhere. Finally, in curve 2c, we can observe all the above mentioned features plus an extra peak located between $E_{Fsample}$ and $E_{Fsample}+\Delta_{sample}$. This peak has been interpreted in terms of an angular anisotropy of the superconducting gap along the Fermi surface of NbSe$_2$[12]. The recent picture of NbSe$_2$ as a multiband superconductor should be taken into account in the interpretation of the angular variation of this peak, and special attention must be addressed to its evolution with temperature.

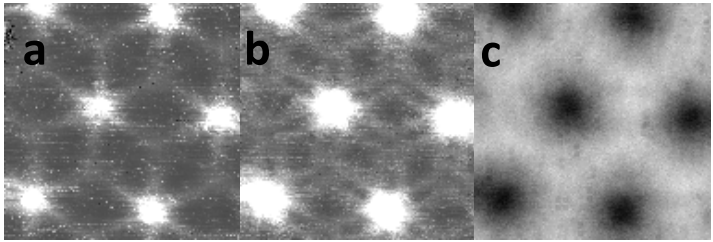

Fig. 3. Normalized conductance images obtained at 300mK and 0.1T corresponding to different energies ((a) 0.8 mV, (b) 1.4mV, and (c) 2mV). Image size: 312x312 nm$^2$.

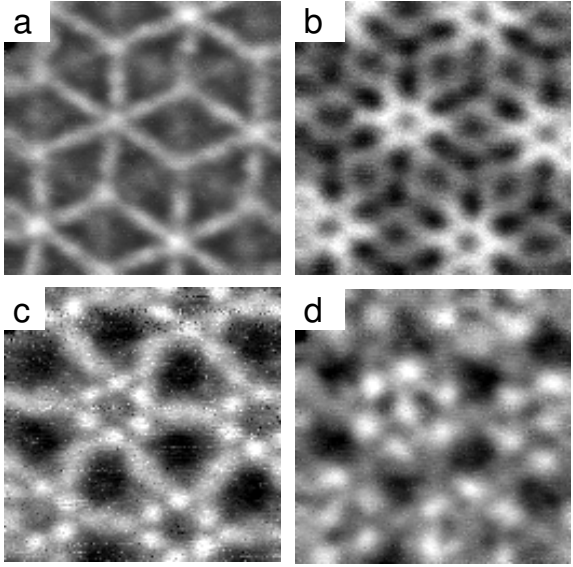 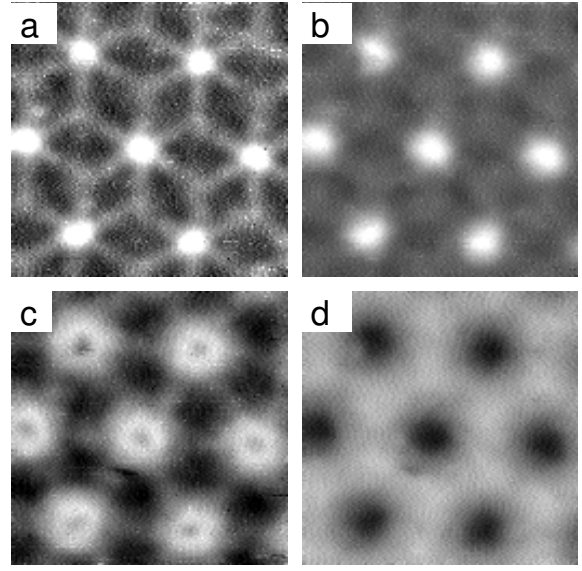

Fig. 4. Conductance ratio images, at 300mK and 1kG, at different energies: (a to d) 0.8, 1.0, 1.2 and 1.5mV. Features around the vortex core at energies between $E_{Fsample}$ and $E_{Fsample} +\Delta_{sample}$ are enhanced, showing also six-fold symmetry. Area size: 312x312 nm$^2$.

Fig. 5. Conductance ratio images, obtained at 6K and 1kG, at different energies: (a to d) 0.8, 1.0, 1.2 and 2.0 mV. The six-fold features around the vortex core previously observed for intermediate energies are now absent. Area size: 364x364 nm$^2$.

Once we have discussed the different features present in the conductance curves, we show in fig.3 three conductance maps corresponding to selected energies between $E_{Fsample}$ and $E_{Fsample} +\Delta_{sample}$. Maps 3a and 3b show the expected six-fold star shaped patterns around each vortex core, due to electronic bound states, whose intensity diminishes as they depart from the core.

The aim of this work is to explore the possibility of observing variations with temperature of the conductance patterns around the vortices, which could add information about multiband character of superconductivity in NbSe$_2$. Therefore, it is necessary to enhance the information contained in the conductance maps, which will become more smeared as temperature is increased.

In order to achieve this enhancement we have produced two methods of spectroscopic imaging, which we name as "Conductance Ratio" and "Conductance Difference" maps. Instead of mapping the conductance, we will map the quantities CR (or CD) corresponding to the ratio (or difference) between the average conductance in two small energy intervals around a selected energy (CR=G1/G2, CD=G1-G2, see fig. 2). These two methods produce a considerable enhancement of local maxima, minima and inflections in the conductance curves, which reflect in the images shown in fig. 4.

This process allows to detect that the "extra" peak indicated in curve 2c appears at different energies and distances to the vortex core, in a six-fold pattern, like the one observed for the bound state at $E_{Fsample}$.

The same imaging methods have been used to produce the images corresponding to the vortex patterns at high temperature. Figure 5 shows CR maps obtained at 6 K. Following the observations and analysis related to MBSC in NbSe$_2$, the superconducting gap should be strongly depressed in one of the bands at this temperature [6]. This situation may be responsible of the absence of the six-fold pattern related to the above mentioned "extra" peak at 6K, while the patterns observed at $E_{Fsample}$ and $E_{Fsample} +\Delta_{sample}$ remain almost identical to those obtained at 300 mK. These results are a new example of the increasing capabilities of the use of a superconducting tip in the field of scanning tunneling spectroscopy.

## 4. Conclusions

As a summary, we have presented STS results on the electronic conductance patterns of a type II superconductor, NbSe$_2$, in the vortex state. These results were obtained using a STM with a superconducting tip. The atomically sharp nature of the fabricated nanotip makes possible the observation of the complex pattern of electronic bound states, with a six-fold symmetry, usually obtained in experiments at very low temperatures with normal tips, as well as a mapping of the Josephson current between tip and sample in this vortex state.

The high spatial and spectroscopic resolution of our superconducting tip allows to obtain information on the conductance patterns even at temperature close to $T_C$. The information contained in these measurements is enhanced by mapping the local conductance ratio or the local conductance difference instead of the usual mapping of the local conductance.

The results obtained from this type of studies of the vortex state in a wide temperature range, will be useful as a new source of information about the nature of multiband superconductivity in NbSe$_2$ and other similar materials.

**Acknowledgements**

Support from ESF-JSPS NES program, Plan Nacional de I+D+I, MEC, Spain (projects Consolider and FIS2004-028977) and the Comunidad Autónoma de Madrid, Spain (program Citecnomik, P-ESP-000337-0505) is acknowledged. We thank I. Guillamón and H. Suderow for useful discussions, as well as C. Dewhurst for providing some of the $NbSe_2$ samples.

**References**

[1]  H. F. Hess *et al.*, Phys. Rev. Lett. **62**, 214 (1989). H.F. Hess et al., Physica B 169, 422 (1991).
[2] S.H. Pan, E.W. Hudson and J.C. Davis, Rev. Sci. Instrum. 70,1459 (1999).
[3] I. Maggio-Aprile et al., Phys. Rev. Lett. 75, 2754 (1995).
[4] J. E. Hoffman et al, Science 295, 466-469 (2002). G. Levy et al., Phys. Rev. Lett. **95,** 257005 (2005)
[5] H. Nishimori *et al.*, J. Phys. Soc. Jpn. **73**, 3247 (2004).
[6] A detailed description of the different investigations and knowledge on this material can be found in the special issue Physica C 385 (2003).
[7] T. Yokoya et al., Science 294, 2518 (2001). E. Boaknin et al. Phys. Rev. Lett. 90, Art. No. 117003 (2003).
[8] J.G. Rodrigo and S.Vieira, Physica C, 404, 306 (2004).
[9]  A. Kohen et al., Phys. Rev. Lett. **97,** 027001 (2006)
[10 ] H. Suderow et al., Phys. Rev. B 65, 100519(R) (2002); J.G. Rodrigo et al., phys. stat. sol. (b) 237, No. 1, 386 (2003); J.G. Rodrigo et al., European Physics Journal B 40 (4): 483 (2004). J.G. Rodrigo et al., J. Phys.: Condens. Matter 16, R1151–R1182 (2004).
[11]  N. Nakai et al., Phys. Rev. Lett. **97,** 147001 (2006).
[12]  N. Hayashi,  M. Ichioka, and K. Machida, Phys. Rev. Lett. **77**, 4074, 1996.